\documentclass{aastex}
\usepackage{emulateapj5}

\slugcomment{Version of 8th May, 2008}

\shorttitle{Fornax dSph I: SFH}
\shortauthors{Coleman \& de Jong}

\begin{document}

\title{A Deep Survey of the Fornax dSph I: Star Formation History}

\author{Matthew G.\ Coleman \& Jelte T.\ A.\ de Jong}
\affil{Max-Planck-Institut f\"{u}r Astronomie, K\"{o}nigstuhl 17, D-69117 Heidelberg, Germany}
\email{coleman@mpia-hd.mpg.de, dejong@mpia-hd.mpg.de}

\begin{abstract}

Based on a deep imaging survey, we present the first homogeneous star formation history (SFH) of the Fornax dwarf spheroidal (dSph) galaxy.  We have obtained two-filter photometry to a depth of $B \sim 23$ over the entire surface of Fornax, the brightest dSph associated with the Milky Way, and derived its SFH using a CMD-fitting technique.  We show that Fornax has produced the most complex star formation and chemical enrichment histories of all the Milky Way dSphs.  This system has supported multiple epochs of star formation.  A significant number of stars were formed in the early Universe, however the most dominant population are the intermediate age stars.  This includes a strong burst of star formation approximately $3-4$ Gyr ago.  Significant population gradients are also evident.  Similar to other dSphs, we have found that recent star formation was concentrated towards the centre of the system.  Furthermore, we show that the central region harboured a faster rate of chemical enrichment than the outer parts of Fornax.  At the centre, the ancient stars (age $> 10$ Gyr) display a mean metallicity of [Fe/H] $\sim -1.4$, with evidence for three peaks in the metallicity distribution.  Overall, enrichment in Fornax has been highly efficient: the most recent star formation burst has produced stars with close to solar metallicity.  Our results support a scenario in which Fornax experienced an early phase of rapid chemical enrichment producing a wide range of abundances.  Star formation gradually decreased until $\sim$4 Gyr ago, when Fornax experienced a sudden burst of strong star formation activity accompanied by substantial chemical enrichment.  Weaker star forming events followed, and we have found tentative evidence for stars with ages less than 100 Myr.

\end{abstract}

\keywords{galaxies: individual (Fornax dSph) --- stars: formation --- Local Group}

\section{Introduction}

The dwarf spheroidal galaxies (dSphs) are the least luminous galaxies known.  They display remarkably high mass-to-light ratios ($\sim$100 to 1000), and the stars in each system are known to reside at the centre of a massive dark halo ($M_{\hbox {\scriptsize vir}} \sim 10^8 - 10^9 M_{\odot}$) which extends far beyond the observed limiting radii \citep{walker07,simon07}.  In terms of stellar population, these systems can have surprisingly complex star formation histories (SFHs).  All dSphs contain a population of ancient stars \citep{held00}, however some (such as Fornax) have been able to maintain multiple epochs of star formation and chemical enrichment over a Hubble time.  These systems are relatively simple environments compared to larger galaxies, and are therefore a starting point in the study of star formation and enrichment.  Simulations suggest a cyclical process, in which the gas collapses to form stars, is then chemically enriched and blown out by pockets of massive star formation, and then collapses again to repeat the cycle.  \citet{salv08} propose a time frame of $\sim$250 Myr for a single cycle.

There are, however, open questions regarding star formation in dwarf galaxies.  Population gradients suggest that the most recent bursts of star formation in each dSph were concentrated towards the object's centre \citep{harbeck01}.  Also, the Milky Way Halo contains a population of extremely metal-poor stars, with abundances of [Fe/H] $< -5$ \citep{christlieb02,frebel05}, whereas the dSphs do not contain a population with metallicities below [Fe/H] $\sim -3$ \citep{helmi06}.  This suggests that the gas sourcing the first generation of stars in these systems was pre-enriched.  Moreover, \citet{grebel04} have argued that the variety of SFHs in dSphs is not the result of reionization, and hence `local processes' have influenced star formation in each object.  These include the regulation of gas dynamics due to internal feedback \citep{dekel86} while ram pressure stripping and tidal interaction with the Milky Way are also important factors \citep{mayer06}.

With an integrated absolute magnitude of $M_V = -13.1$ \citep{m98}, Fornax is the brightest dSph associated with the Milky Way (excluding the tidally dirsupting Sagittarius system).  It lies at a distance of 138 kpc, and proper motion measurements indicate its orbit around the Milky Way is roughly circular with an eccentricity of $e=0.13^{+0.25}_{-0.02}$ and an orbital period of $3.2^{+1.4}_{-0.7}$ Gyr \citep{piatek07}.  \citet{walker06} have completed a large kinematic survey of this system, collecting radial velocities for 206 member stars over the entire surface of Fornax.  They measured a flat velocity dispersion profile, indicating a significant dark component, and find a mass-to-light ratio of $M/L_V \sim 15$ within 1.5 kpc (approximately half the tidal radius).

Compared to other dSphs, the star formation history of Fornax is unusually complex.  All dSphs contain some number of ancient stars \citep{grebel04}, however early studies of Fornax revealed an extended giant branch, including carbon stars, suggesting a strong intermediate-age component \citep{demers79,aaronson80,aaronson85}.  Additionally, Fornax contains a significant young stellar component.  Originally discovered by \citet{beauchamp95}, the photometry of \citet{stetson98} and \citet{saviane00} later identified main sequence stars with ages as young as 100--200 Myr.  The analysis of \citet{gallart05} indicates a burst of star formation in the centre of Fornax 1--2 Gyr ago, which has continued almost to the present day.  Indeed, more than half the stars on the RGB are thought to be younger than 4 Gyr \citep{batt06}.  Of all the dSphs, Fornax has experienced the most recent star formation.

Fornax also contains population gradients, in which the young stars are preferentially located towards the centre \citep{stetson98,batt06}.  This property is common in the dSph population \citep{harbeck01} and suggests that the gas required for subsequent star formation episodes was more successfully retained in the core of the dark halo than in the outer regions.  In Fornax, the young component is not aligned with the main body and is highly structured, including a shell-like feature which may indicate an accretion event $\sim$2 Gyr ago \citep{coleman04,ols06}.

In addition to an age spread, the stars in this system cover a significant range in metallicity.  \citet{tolstoy01}, \citet{pont04} and \citet{batt06} have examined the chemical abundances of Fornax red giants using spectra of the Ca II triplet features, and found a metallicity range of $-2.5 \le$ [Fe/H] $\le 0.0$.  The data set of Battaglia et al.\ contained enough Fornax members (562) such that stellar metallicities could be accurately related to kinematics.  They found that the metal-rich stars have a colder velocity dispersion, and the metal-poor component shows signs of non-equilibrium kinematics towards the centre of Fornax ($r < 2r_c$).  Additionally, high resolution spectra of 81 red giants in the centre of Fornax indicate that s-process elements are unusually strong, hence stellar winds (such as those from AGB stars) have dominated the chemical enrichment of Fornax in the last 2--4 Gyr \citep{letarte07}.

Despite the progress described above, Fornax has been lacking a homogeneous study of its SFH over the entire surface of the system.  The {\em HST} photometry of \citet{buon99} confirmed the presence of young stars at the centre of Fornax and included evidence of seperate bursts of star formation.  By combining these results with Ca II triplet metallicities, \citet{tolstoy01} created {\em schematic} star formation and chemical enrichment histories.  The preliminary SFH produced by \citet{gallart05} was based on VLT/FORS1 photometry (depth $I \sim 24.5$) also located at the centre of Fornax, and another field approximately one core radius from the centre.  Young stars were present in both fields, however the authors noted a significant difference in SFH between the two.  In summary, although the general trend of star formation at the {\em centre} of Fornax is known, the aggregate history is yet to be determined.

Hence, we present the first results of a deep, homogeneous photometric data set over the entire surface of Fornax.  We have extracted the SFH from this photometry using the CMD-fitting techniques developed by \citet{dolphin02}.  Not only has this allowed us to derive a global SFH for this system, we have also examined the SFH as a function of position to search for population gradients.  Our data have a limiting magnitude of $B \sim 23.0$, thus we are sensitive to main sequence stars in Fornax with an age of 3 Gyr and less.  Also, the red giant branch, red clump and horizontal branch stars allowed us to track the ages and metallicities of the populations with ages $>3$ Gyr.  This is the first complete SFH of this system derived from deep photometry.

\section{The Survey}

\subsection{Data Reduction}

Images encompassing the surface of the Fornax dSph were obtained using the ESO/MPG 2.2m Telescope equipped with the Wide-Field Imager.  This instrument provides a $34' \times 33'$ field of view using a $4 \times 2$ mosaic of $2048 \times 4096$ pix${}^2$ CCDs and a pixel resolution of $0.24''$ pix${}^{-1}$.  The survey aim is to obtain photometry to 23rd magnitude over the entire body of Fornax.  The first stage presented here contains Fornax itself and the outer shell noted by \citet{coleman05}.  Thus far, 21 pointings have been obtained, covering a sky area of $5.25$ deg${}^2$.  An overlap region of $\sim$$4'$ between each field was chosen to ensure the photometric zeropoint was constant across the survey.  A schematic diagram of the fields is shown in Fig.\ \ref{obsmap}.  We obtained three 600s dithered exposures in both $B$ and $R$ for all fields, allowing us to reach magnitudes of $B = 23.0$ and $R = 23.5$ ($50\%$ completeness limits) in all fields.  The images were taken during 11 nights in Oct/Nov 2006 in median seeing conditions of $1.4''$ (range $0.9'' - 2.1''$).  The data were reduced using standard procedures in the {\em mscred} package in IRAF: the overscan region and the bias frames were used to subtract the pedastal current from each science image, which were then trimmed.  Twilight flat fields were combined to produce a master flat frame in $B$ and $R$ for each night, which were then used to flat-field the science images.

\begin{center}
\plotone{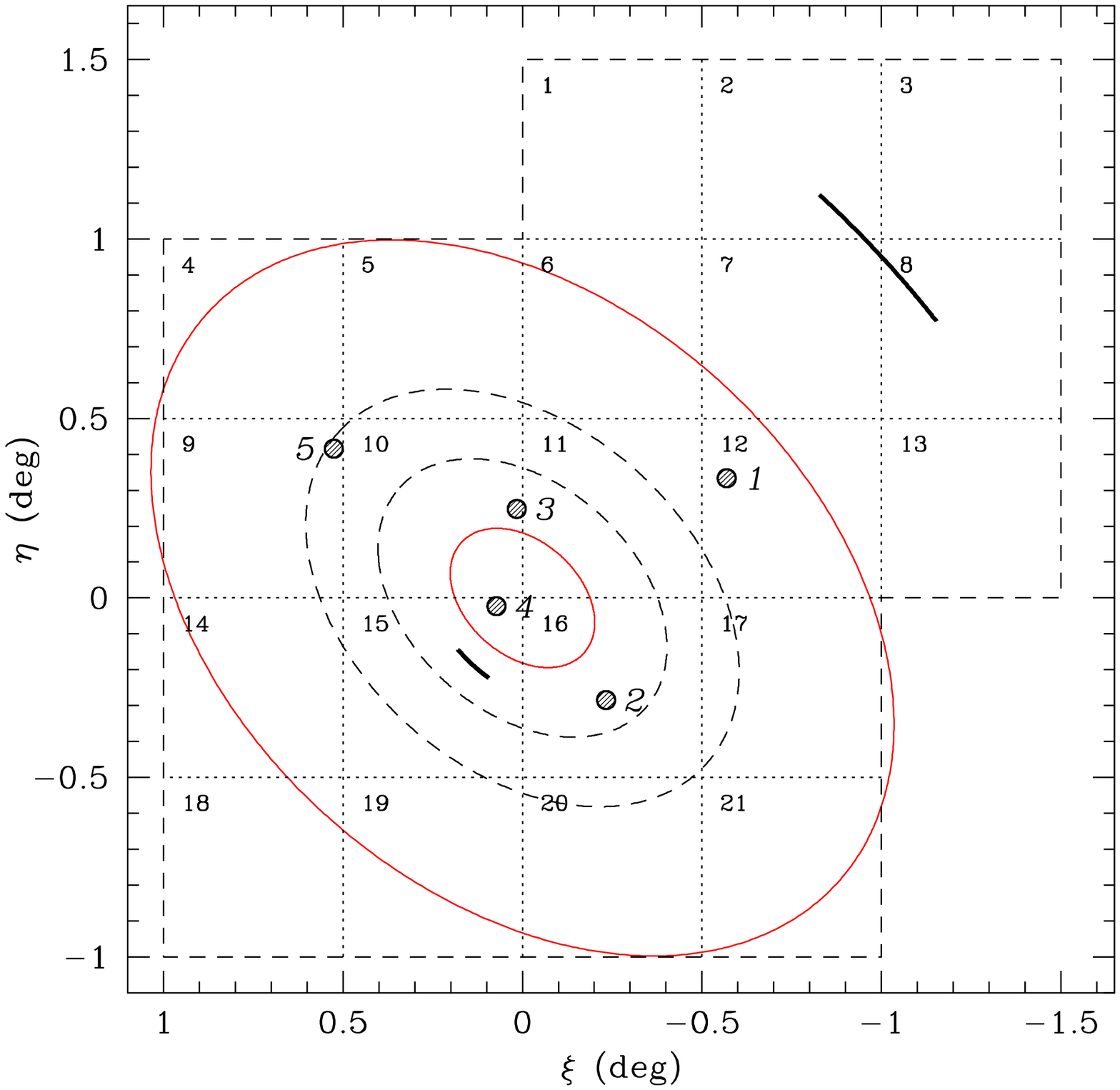}
\figcaption[Fornax observation map]{Schematic diagram detailing the layout of the fields currently observed.  The red ellipses represent the core and tidal radii \citep{m98}.  The dashed lines are the limit of our survey and the dotted lines represent the overlap regions between fields.  The shaded circles represent the five globular clusters associated with Fornax and are labelled accordingly.  These circles have radii of $1.5'$, approximately equal to the tidal radii of each cluster \citep{mackey03}.  The two arcs represent the shell-like features noted by \citet{coleman04,coleman05}, and the dashed ellipses represent the boundaries of our radial bins.  \label{obsmap}}
\end{center}

Three dithered images were taken in each field, and we used routines in the {\em mscred} package to combine the images and remove gaps between the CCDs \citep{valdes02}.  An astrometric solution was constructed for all images by matching them to the first USNO CCD Astrograph Catalogue (UCAC1), which has an average precision of 31 mas in the magnitude range $8 < R < 16$ \citep{zach00}.  The {\it rms} of our solution was $<$$0.2''$ in all fields.  The individual CCD images were then combined using the {\em mscimage} routine to produce three single (i.e.\ non-mosaic) images for each field in both filters.  These three images were then median combined with the {\em mscstack} routine, which matches images based on their astrometry and removes the CCD gaps.  Finally, the combined image for each field was then corrected for zeropoint gradients across the field of view using the {\em mscskysub} routine.

\subsection{Photometry}

Photometry was derived using DAOPHOT \citep{stetson87}.  We measured the background level in each field to calculate a standard deviation of the sky, $\sigma$.  Each image was then searched for all sources $4\sigma$ above the background level, and aperture photometry was used to estimate their brightnesses.  In a crowded field such as the centre of Fornax, the PSF-fitting technique in DAOPHOT provides a more accurate measure of stellar magnitudes compared to aperture photometry, as it allows the signal of adjacent stars to be disentangled.  Hence, we examined the brightest 60 stars in each image and used those with no apparent neighbours and a well-defined Gaussian shape to construct a master PSF for each science frame.  This was then fitted to every source in the image using a fitting radius of 8 pixels ($1.9''$) or $1-2$ half-width, half-maxima of the PSF (depending on seeing).

To measure the completeness and photometric accuracy of our photometry, we performed artificial star tests on all science frames in both filters (i.e.\ 42 science images).  We placed 1600 artificial stars in the image and attempted to recover them with DAOPHOT, where the photometric uncertainty was then determined as the dispersion of the returned magnitudes about the mean (that is, not the input).  This was repeated for artificial stars at every 0.25 magnitudes in the $B$ and $R$ frames.

To ensure a constant zeropoint across the survey, we matched stars in the overlap regions between fields using their astrometry and measured the mean inter-field difference in $B$ and $R$.  This is the same technique used in our previous Fornax survey \citep{coleman05}.  The inter-field corrections were accurate to $\sim$$0.02$ mag in both $B$ and $R$.  As an example, the final match between the fields F6 and F11 is shown in Fig.\ \ref{f11f6}.  A final zeropoint correction was made by matching our data to the catalogue of \citet{stetson98}, which contains photometry of the core of Fornax in $B$ and $R$ to a similar limiting magnitude as our survey.  The $R$ filter attached to the 2.2m/WFI is a standard Cousins filter, and the $R$-band photometry was well matched between the two datasets.  In contrast, the $B$ filter (BB\#B/123\_ESO878) covers a larger wavelength range than the standard Johnson filter\footnote{http://www.ls.eso.org/lasilla/sciops/2p2/E2p2M/WFI/filters/}, and therefore requires a colour correction to match standard $B$ magnitudes.  The ESO website provides a correction for the $B$-band photometry using the $(B-V)$ colour; we determined a $(B-R)$ colour correction by comparing our data to the Stetson et al.\ catalogue, yielding the following result,
\begin{displaymath}
B = b +0.20(b - R),
\end{displaymath}
with a bootstrap error of 0.02 mag.  The overall photometric zeropoints are accurate to 0.03 mag.

\begin{center}
\plotone{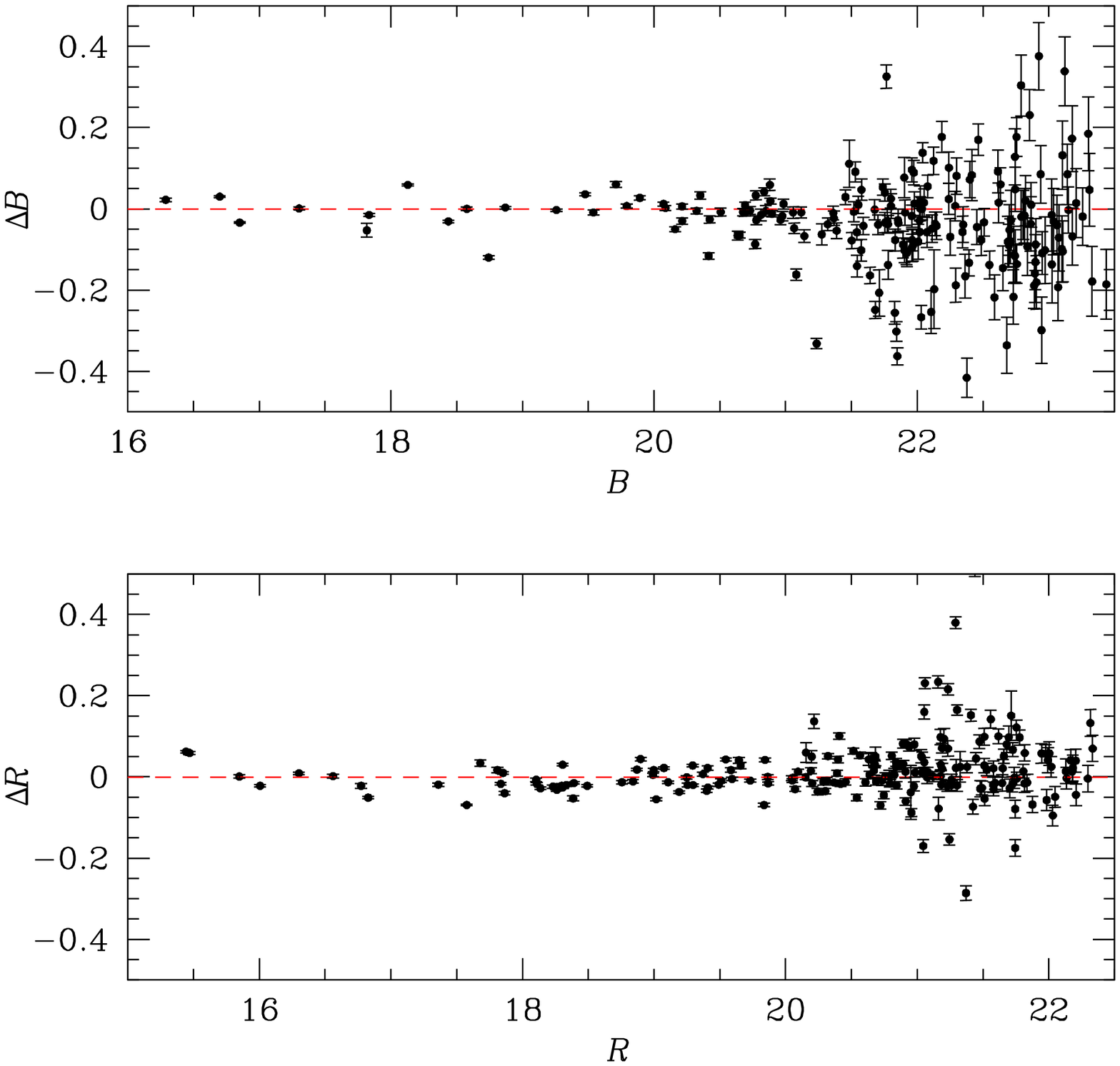}
\figcaption[Field comparison]{A comparison of the photometry measured in fields 6 and 11.  We have removed all sources with a non-stellar sharpness and large photometric uncertainty ($\sigma_{(B-R)} > 0.1$ mag) \label{f11f6}}
\end{center}

\subsection{Colour Magnitude Diagram}
The colour magnitude diagram (CMD) of Fornax reinforces the complex star formation history (SFH) of this object: it contains multiple stellar populations with a vast range in age and chemical abundance.  A full description of the stellar populations at the centre of Fornax is given by \citet{stetson98} and \citet{saviane00}.  However, the outer regions of Fornax are not well known, and in Fig.\ \ref{cmd} we present the first deep CMDs for the entirety of Fornax.  It has been theorised that Fornax may contain strong radial population gradients (e.g.\ \citealt{saviane00}), hence we have divided the data set into the four elliptical regions shown in Fig.\ \ref{obsmap}.  The regional boundaries are at radii of $r_c$, $2r_c$, $3r_c$ and $r_t$, where $r_c = 13.8'$ and $r_t = 76.0'$ are the core and tidal radii listed by \citet{m98}.  The distribution of young stars in Fornax is not aligned with the system's major axis and it is also known to contain strong asymmetries (e.g.\ \citealt{stetson98}), however the majority of these stars are contained in the core region (this will be shown in the next paper in this series; Coleman, in prep.) and hence the assumption of elliptical regions is not vital for this sub-population.

\begin{center}
\plotone{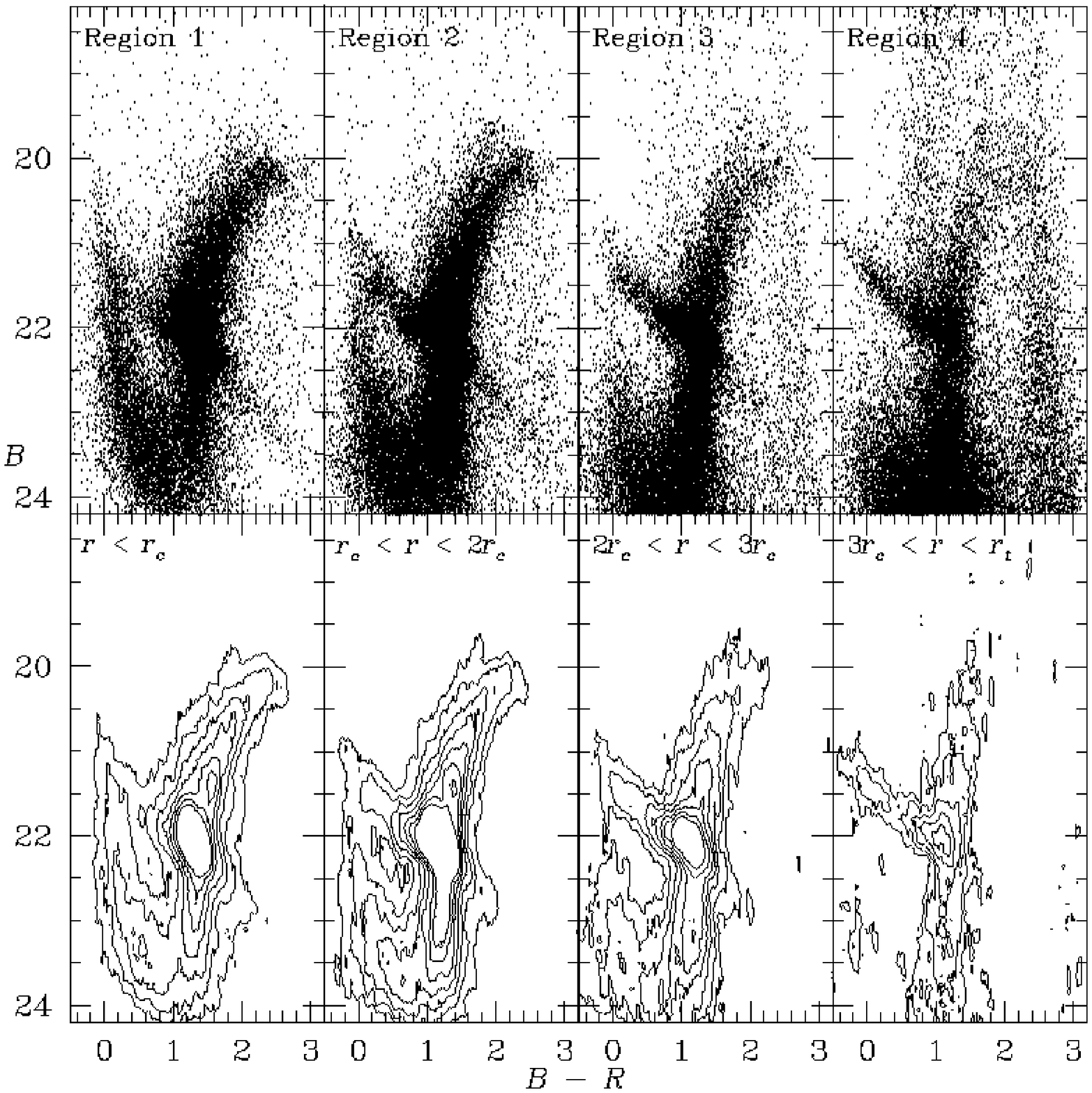}
\figcaption[Fornax CMD]{Colour-magnitude diagrams of the four Fornax regions.  The upper four panels show the photometry for sources in Region 1 ($r < r_c$), Region 2 ($r_c < r < 2r_c$), Region 3 ($2r_c < r < 3r_c$) and Region 4 ($3r_c < r < r_t$).  The photometry is $95\%$ complete to a magnitude of $B=23.0$, however this number improves slightly ($\sim$0.25 mag) in the outer regions due to less stellar crowding.  We have removed all sources with a non-stellar sharpness and a large photometric uncertainty ($\sigma_{(B-R)} > 0.3$ mag).  The four lower panels represent the `signal' of each subset above the background.  \label{cmd}}
\end{center}

Region 1 (the core region) contains approximately 28,000 sources.  Moving outwards, another 41,000 stars were selected in Region 2, 31,000 in Region 3, and 51,000 in Region 4.  For convenience, we have also performed a background subtraction on each CMD, removing the field population using the `signal-to-noise' technique provided by \citet{grillmair95}.  A full description of this technique as applied to the current data set will be given in the next paper in this series.  In summary, we have divided the CMD into a grid of cells, counted the number of Fornax and field stars in each cell, and then used this to remove the field contamination (mostly foreground Milky Way stars).  The results are shown in the lower panels of Fig.\ \ref{cmd}.

The central CMD (upper left panel) shows all sources in Region 1, drawn from within the inner red ellipse shown in Fig.\ \ref{obsmap}.  Immediately visible is the red giant branch extending downwards from $B \sim 20$, which contains old and intermediate age stars (age $> 1$ Gyr) and is thickened due to the age and metallicity spread.  The most densely packed feature is the red clump (RC), centred at $(B-R)=1.3$, $B=22$.  This is made up of core helium burning stars, and is essentially a young-to-intermediate age, metal-rich horizontal branch (HB).  A hint of a HB extending towards the blue is also visible.  We also note that the HB appears to extend redward from the RC, however these objects are artefacts of the observational dithering pattern: a CMD-selection indicated that they lie pre-dominantly in regions in which only a single frame of $B$-band data was available.  This effect is most prominent in the HB star-rich central four fields, where a few of the HB stars display large photometry errors in the $B$ filter (and hence an artificial colour spread in the HB itself is created).  Although it is a small artefact (it exists in less than $1\%$ of the surveyed area), its effect on our star formation history results is discussed further in \S \ref{sec:abvar}.

Returning to our discussion of the central region CMD, a slight overdensity lies approximately $0.6$ mag above the RC, identified as the colour-magitude clumping at the start of the AGB cycle \citep{saviane00}.  Below the RC lies the sub-giant branch, which is also thickened by the age and metallicity spread.  Finally, the blue column of stars extending downwards from $(B-R)=-0.2$, $B=20$ is the young main sequence.  \citet{beauchamp95} discovered this feature in Fornax, and \citet{saviane00} subsequently identified main sequence stars as young as 200 Myr.  Indeed, of all the Milky Way dSphs, Fornax has the most recent star formation.

At first glance, the thick red giant and sub-giant branches are common to all four regions, indicating that the wide range of stellar ages and metallicities continues well beyond the core radius.  However, the differences encountered in the stellar population when moving outwards from the centre of Fornax are remarkable.  We see a decrease in the prominence of the young main sequence, possibly indicating that recent star formation (i.e.\ less than 4 Gyr ago) was preferentially located towards the centre of Fornax.  Another clear difference between the four regions lies in the morphology of the HB.  The red clump is present in all four regions, however the HB extends further into the blue region as we move outwards; it terminates at $(B-R) \sim -0.2$ in the Region 4 CMD.  This indicates that the outer regions of Fornax contain a significant (if not dominant) population of old, metal-poor stars.  Further emphasis of this point is provided by the red giant branch (RGB): the lower panels of Fig.\ \ref{cmd} show that as we move outwards, the RGB shifts towards the blue, thus indicating a decrease in mean metallicity with increasing radius.  Overall, a comparison of the four CMDs shown in Fig.\ \ref{cmd} would suggest that later bursts of star formation and chemical enrichment were preferentially located towards the centre of Fornax.
  
\section{Star Formation History}

Numerical fitting of CMDs allows a study of the SFH of a dwarf galaxy \citep[e.g.][]{gallart96, tolstoy96, aparicio97, dolphin97, holtzman99, olsen99, hernandez00, harris01}.  Moreover, the large mosaic presented in this paper provides an opportunity to study the spatial variation of the SFH within Fornax.  To obtain a detailed picture of the SFH we use the CMD-fitting software MATCH \citep{dolphin02}, which applies maximum-likelihood methods to fit photometric data with simple model CMDs.  By converting data and models to so-called Hess-diagrams (2-D histograms of the stellar density as a function of colour and magnitude; \citealt{hess24}) a direct pixel-by-pixel comparison is possible.  The model CMDs are based on theoretical isochrones from \cite{girardi02} and include realistic photometric errors and completeness, which are obtained from the artificial star tests described earlier.  By determining the best-fitting linear combination of model CMDs for different age and metallicity bins, the SFH and metallicity evolution are then constrained.  The accuracy of the recovered metallicities depends not only on the data quality, but also on the quality of the isochrones and the stellar evolution tracks on which they are based.  \cite{sdssmatch} tested MATCH on a set of six globular clusters with varying metallicities using isochrones based on the same stellar evolution tracks from \cite{girardi02}.  They show that the recovered metallicities are always within 0.2 dex of the spectroscopic values.  In all results presented in this paper we therefore include an additional contribution to the metallicity uncertainties of 0.2 dex.

Since there are variations in seeing and sky brightness between the different fields in the mosaic, the SFH fits are done separately for each field.  Furthermore, the four radial bins are treated separately to enable an analysis of the radial variation of the stellar populations in Fornax.

\subsection{CMD fitting method}
\label{CMDfit}

The main free parameters in CMD fitting are distance, age, metallicity and extinction, although the binary fraction and the assumed initial mass function (IMF) also play a role.  To limit the number of free parameters to the age and metallicity, we assume reasonable priors on the others parameters.  For all our fits we assume a binary fraction of 0.5 and a Salpeter IMF \citep{salpeter}, which to the CMD depth probed here is practically equal to, for example, a Kroupa IMF \citep{kroupa93}.  During the past decade, several different studies have all found consistent distances to Fornax of $138 \pm 4$ kpc \citep{m98,bersier00,rizzi07}, justifying a prior on the distance in our SFH fits.  To account for the small uncertainty in the distance, we perform all fits for three fixed distances, namely 135, 138 and 141 kpc.  According to the dust extinction maps from \cite{sfd} the foreground reddening towards Fornax varies between $E(B-V) = 0.015$ and $0.03$ mag.  Fits are performed for different combinations of foreground and internal extinctions, with the former fixed at $E(B-V) = 0.015$, 0.02, 0.025 and 0.03 mag, and the latter at $E(B-V) = 0.0$, 0.1, 0.2, 0.3 and 0.4 mag.

For each individual fit, the distance and two kinds of extinction are fixed, while the star formation rates (SFRs) for the age and metallicity bins serve as the free parameters.  Since the isochrones are spaced more evenly in log($t$) than in $t$, the age bins are defined in log($t$).  They have bin-widths of $\Delta \log{t} = 0.15$, with the oldest bin corresponding to ages between 11 and 16 Gyr and the youngest to 10 and 16 Myr, for a total of 21 age bins.  The full metallicity range spanned by the isochrones, [Fe/H] $= -2.4$ to 0.0 is covered with 16 bins of 0.15 dex width.  To account for contamination by foreground stars and faint background galaxies, a control field CMD is created from all star-like sources outside the limiting radius of Fornax, and used as an additional model CMD in the fits.  All stars with $-1 < B-R < 3$ and $B < 23$ and $R < 23$ are fit, using Hess-diagram bin sizes of 0.16 in magnitude and 0.08 in colour.

\begin{center}
\plotone{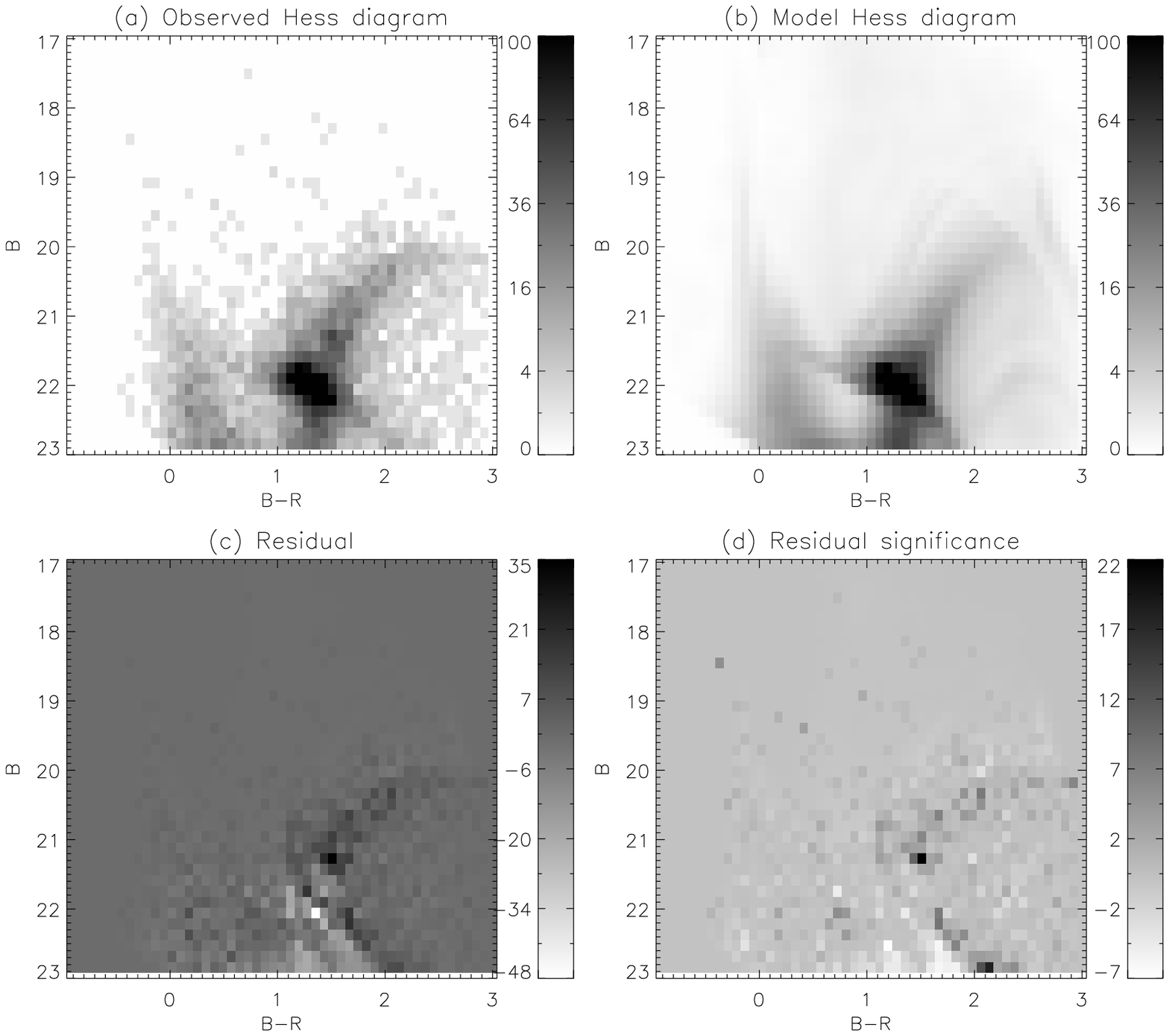}
\figcaption{CMD fit for the central radial bin part of survey field 15.  {\it (a)} Observed Hess diagram, with the grayscale indicating the number of stars per pixel.  {\it (b)} Hess diagram of the best-fit model, including theoretical models and a control field population.  {\it (c)} Residuals after subtracting the model from the data, where darker pixels correspond to underprediction and lighter to overprediction of stars by the model.  {\it (d)} Residuals scaled by Poisson $\sigma$.  \label{fig:f15_bin1_fit}}
\end{center}

Figs.\ \ref{fig:f15_bin1_fit} and \ref{fig:f15_bin3_fit} show the Hess diagrams of Regions 1 and 3 recovered from Field 15 with the corresponding best-fit models and residuals.  In general the fits are good, but some systematic problems arise in reproducing the exact shape of the HB/RC and its extent towards the red.  This imperfect modeling of the HB/RC region is a general problem for theoretical isochrones, due to the complicated processes taking place during this phase in stellar evolution.  Newer isochrone sets than the one currrently used by MATCH \citep{girardi02} should improve this situation, but our current results are not strongly affected by these problems.  However, as the details of the theoretical isochrones used do influence the exact values of metallicities and ages we recover, some extra uncertainty should be taken into account when interpreting the fit results.

\begin{center}
\plotone{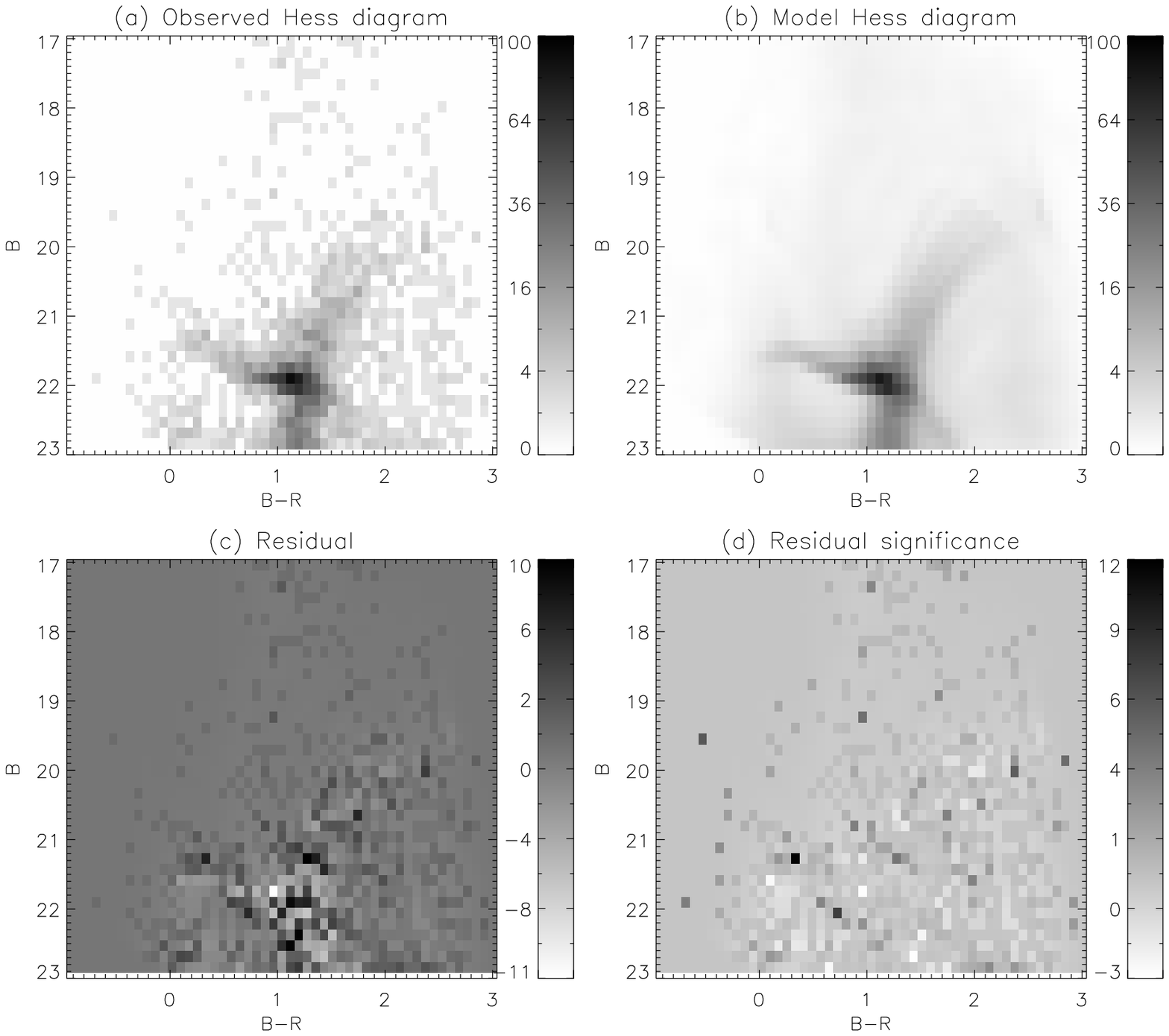}
\figcaption{ Same as Fig.\  \ref{fig:f15_bin1_fit}, but for the third radial bin.  \label{fig:f15_bin3_fit}}
\end{center}

After running the 60 fits (3 distances, 4 foreground and 5 internal extinction values) for each region, the values of the maximum-likelihood measure of goodness-of-fit $Q$ \citep[see][]{dolphin02,sdssmatch} of all fits are compared.  The expected random (i.e.\ not due to actual SFH differences) variance in this $Q$ parameter, $\sigma$, when fitting a specific CMD, is calculated using Poisson statistics and from Monte Carlo simulations using random drawings from the best model.  All fits that have a value of $Q$ within 1$\sigma$ of the best fit are considered as `good' fits and used to construct the SFHs presented in the remainder of this paper.

\subsection{Results}

\subsubsection{A Global View}

As a first pass, we examine the aggregate SFH of the Fornax system.  Combining the SFHs obtained from all regions and fields gives the total SFH of Fornax, presented graphically in Fig.\ \ref{fig:totalsfh} and in tabular form in Table \ref{tab:totalsfh}.  Clearly, Fornax has a complex history and has been forming stars continuously over most of the age of the universe, with non-zero SFRs being found in most age bins.  After the first stars formed more than 10 Gyr ago, a slow decline is detected until approximately 4 Gyr ago, when the star formation rate sharply increased for a period of $\sim$1 Gyr.  After this sudden intensified star formation episode, a very low level of star formation has continued until the present day.  For comparison, we also show in Fig.\ \ref{fig:totalsfh} the schematic SFH derived by \citet{tolstoy01} based on photometry and Ca {\sc ii} triplet results.  We have shifted the peak of their relative star formation rate to match our peak SFR.  The results are well matched for the last few Gyr, however our CMD-fitting code is better able to extract detailed star formation rates from the early history of Fornax.  We also show the total stellar mass formed during each bin in Table \ref{tab:totalsfh} and Fig.\ \ref{masshistory}.  From this, we calculate the total stellar mass formed in Fornax to be $6.1^{+0.8}_{-0.7} \times 10^7$ $M_{\odot}$, where we have integrated the Salpeter IMF down to a mass of $0.15 M_{\odot}$.

\begin{center}
\plotone{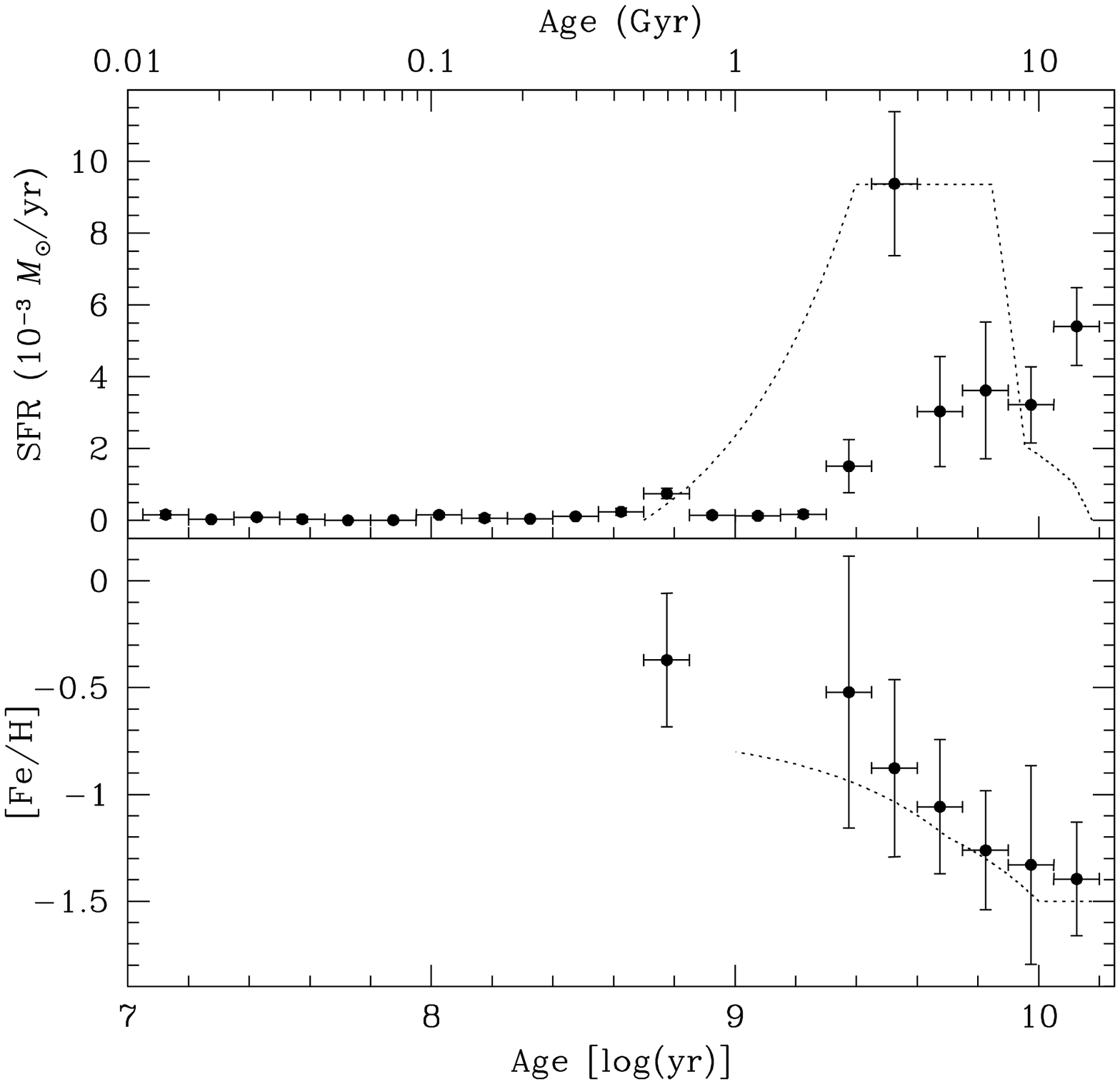}
\figcaption{Total star formation history ({\it upper panel}) and age-metallicity relation ({\it lower panel}) for Fornax.  In both panels, the horizontal error bars indicate the widths of the age bins.  The error bars on the star formation rates are not independent; a higher SFR in one bin would be compensated by a higher SFR in adjacent bins.  Metallicities are only shown for age bins with a SFR greater than 0.0005 $M_\odot~ yr^{-1}$ and are SFR-weighted means.  Both panels contain a comparison with previous work, represented by the dotted lines.  The upper panel contains the schematic SFH derived by \citet{tolstoy01} scaled to match our peak star formation rate detection, and the lower panel contains our estimate of the evolution of the average metallicity derived by \citet{batt06}.  \label{fig:totalsfh}}
\end{center}

\begin{center}
\plotone{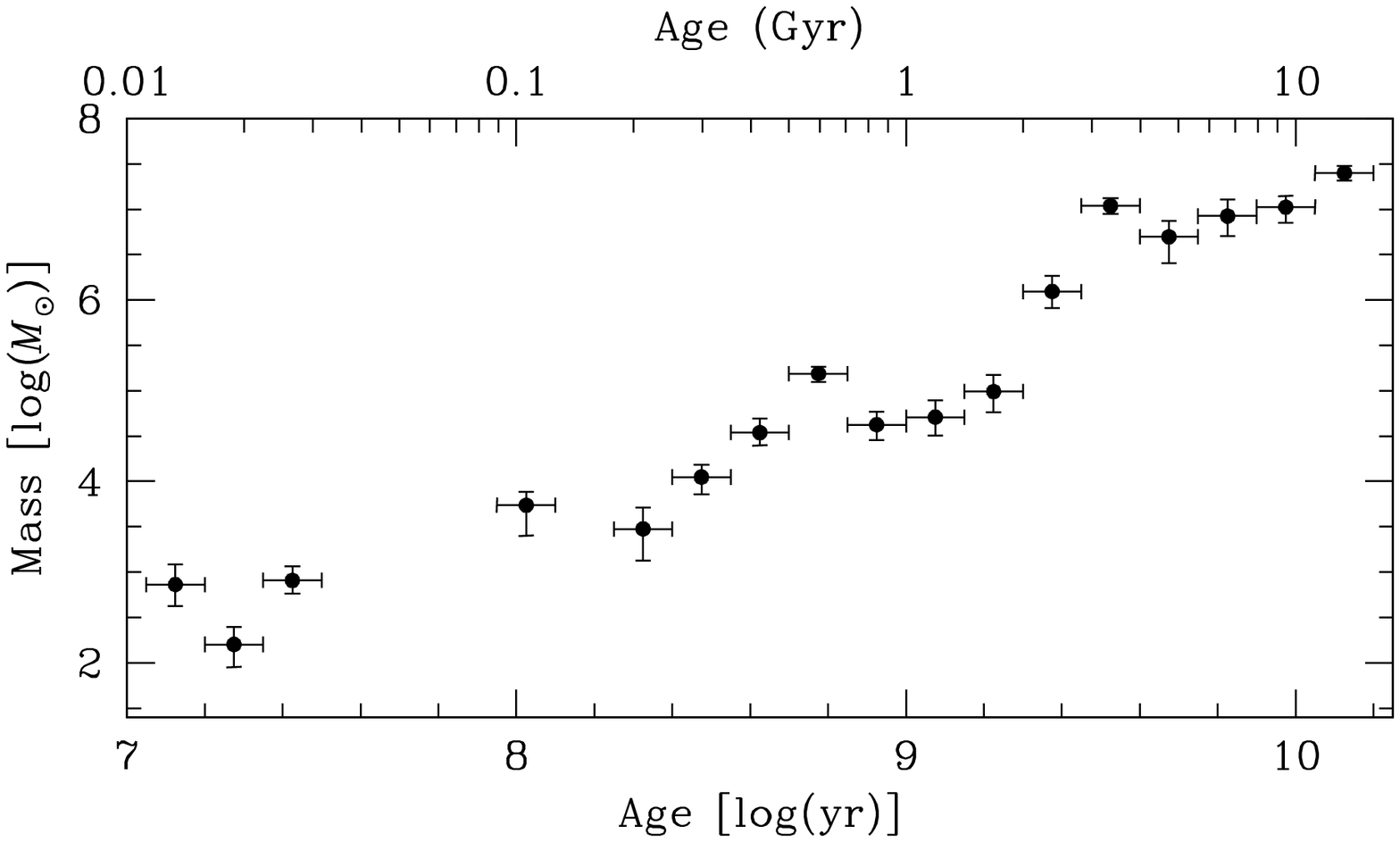}
\figcaption{Stellar mass formed during each age bin, taken directly from the total star formation history (Fig.\ \ref{fig:totalsfh}).  \label{masshistory}}
\end{center}

Initially, the average metallicity of the stars shows little evolution.  A large fraction of the ancient stars seems to have formed from pre-enriched gas, as the mean metallicity in the oldest age bin ($11-16$ Gyr) is already [Fe/H] $= -1.4$.  The metallicity only starts to increase significantly from this value at an age of around 5 Gyr.  Overall, these results are in good agreement with previous studies of the SFH and metallicity of Fornax.  We show in the lower panel of Fig.\ \ref{fig:totalsfh} the approximate age-metallicity relation derived by \citet{batt06} based on spectra of 562 RGB stars extending from the centre of Fornax to the tidal radius.  This line represents our estimate of the evolution of the average iron abundance from Fig.\ 23 of \citet{batt06}, however we note that each age bin encompasses a wide range of stellar metallicities and thus this line should be regarded as a guide only.  Although they are well within the uncertainties, our results are slightly more metal-rich than the spectroscopic results.  Such a small offset in metallicity might be caused by slight inaccuracies in the isochrones, the photometric calibration, or the distance to Fornax.

\subsubsection{A Refined View}

Population gradients are a common feature of dSphs \citep{harbeck01}.  Generally, more recent star formation events are more centrally concentreated and produce stars with a greater metallicity compared to the older stellar population.  In Fig.\ \ref{fig:radialsfh} we show the total SFHs (left) and metallicity evolutions (right) for all four radial regions (e.g.\ for Region 1 the four SFHs of the portions in fields 10, 11, 15, and 16 have been combined).  Overplotted in each panel are the SFHs and metallicities for four individual regions, to give an idea of the field-to-field variation.  Since the panels on the right only show the {\em average} metallicities and provide no insight into the metallicity distribution, the combined fit results for the four radial regions with the complete age-metallicity grid are shown in Fig.\ \ref{fig:radialfit}.  

\begin{center}
\plotone{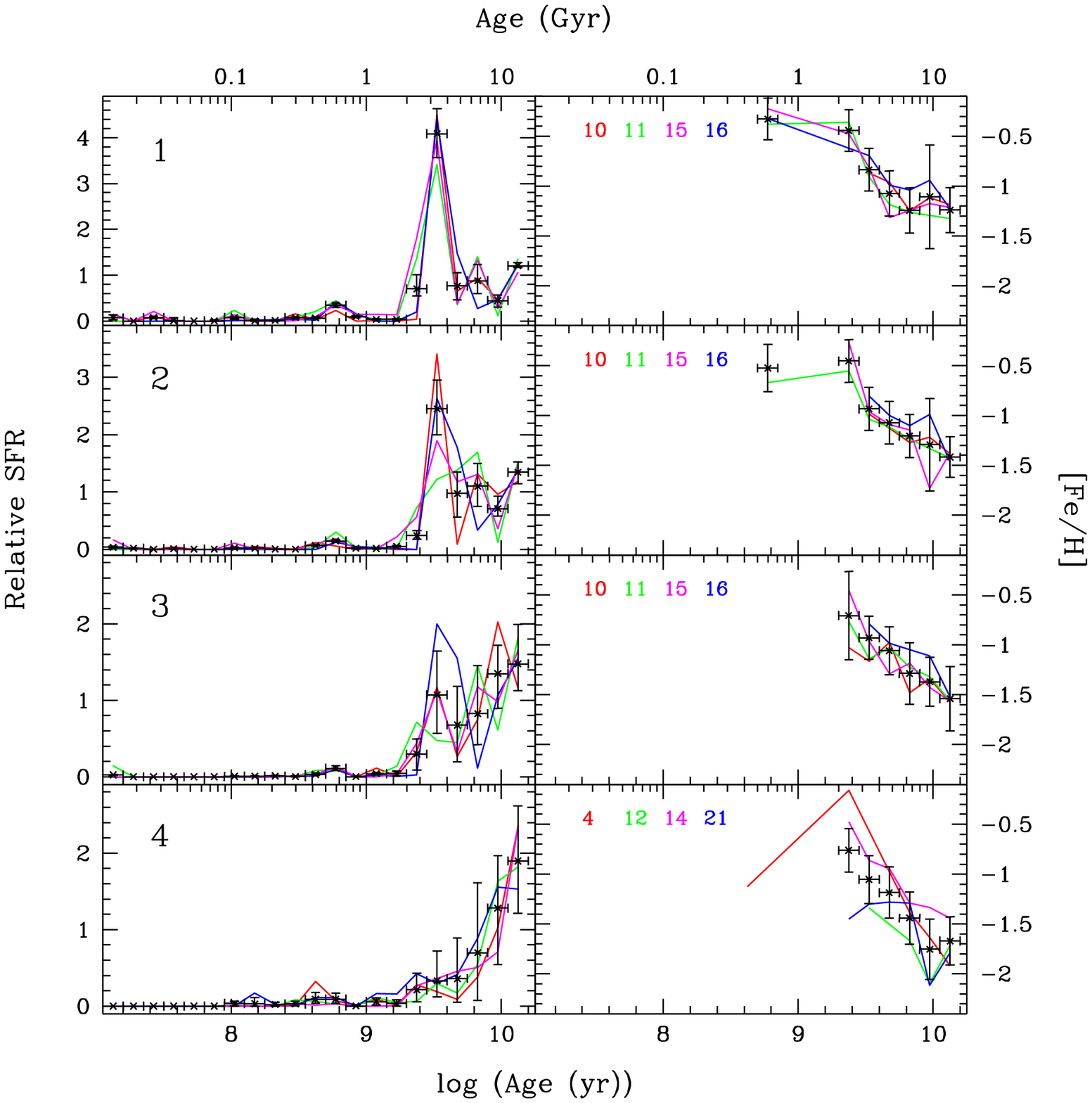}
\figcaption{ Relative star formation histories (SFH,{\it left}) and age-metallicity relations (AMR, {\it right}) per radial bin.  From top to bottom the panels correspond to the central through to the outermost radial bin.  As in Fig.\  \ref{fig:totalsfh}, horizontal error bars show the age bin widths and the errors on the SFRs of adjacent bins are not independent.  The SFHs shown here are scaled to the average SFR in each radial bin.  Overplotted on each panel, using colored lines, are the SFHs and AMRs of each radial bin in four individual survey fields; which fields are plotted is indicated in the panels on the right.
\label{fig:radialsfh}}
\end{center}

There is a striking difference in the SFH when moving from Region 1 (top panels) to Region 4 (bottom panels).  In the center of Fornax, the burst of star formation that occurred $3-4$ Gyr ago stands out strongly.  Moving outwards, the proportion of stars produced during this burst decreases and in the outskirts are not found at all.  Stars younger than 1 Gyr are also mostly found in the center, with hardly any significant star formation at these ages in the outermost radial bin.  This kind of radial dependence, with young stars more centrally concentrated than old stars is a common characteristic of dSphs, and evidence for this was found before in Fornax \citep{stetson98,batt06}.  At all radii, the metallicity evolution follows the same pattern.  It starts off at [Fe/H] $\sim -1.5$ and remains constant until $\sim$5 Gyr ago.  Then the metallicity starts to increase rapidly to $-0.5$, after which there is little indication of further enrichment.  This rapid increase in metallicity coincides with the strong burst of star formation.

\begin{center}
\plotone{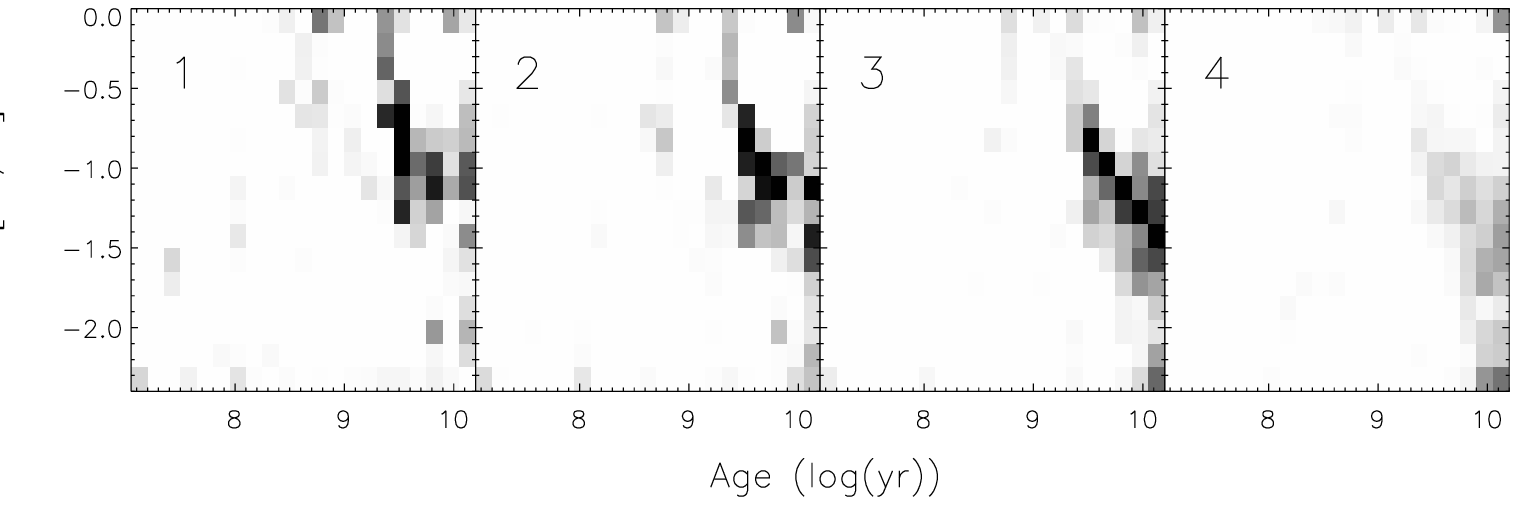}
\figcaption{ CMD fitting results for the four radial bins.  From left to right the panels are for the central bin (labeled with a ``1'') to the outermost bin (labeled with a ``4'').  The grayscale indicates the star-formation rate in each age-metallicity bin, with a darker value corresponding to a higher value.  \label{fig:radialfit}}
\end{center}

Comparing the SFHs of individual fields within a radial bin (i.e.\ the coloured lines in Fig.\ \ref{fig:radialsfh}) shows no significant difference in the central bin.  Since in the central part the dynamical timescales are shortest, the populations are well-mixed.  In the outermost region there is also no sign of spatial variation, which is probably because there are mostly ancient stars which have had sufficient time to diffuse throughout the galaxy.  In Regions 2 and 3, however, significant field-to-field variation is visible in the strength of the 4 Gyr burst.  The detailed structural properties of the stellar populations in Fornax is the subject of the next paper in this series (Coleman, in prep.).

\subsubsection{Abundance Variations in the Ancient Stars} \label{sec:abvar}

Although Fornax is found to be relatively metal-rich, Fig.\ \ref{fig:radialfit} shows that the spread in metallicities is significant.  At all ages the metallicity spread is at least $\sim$1 dex, and especially the oldest stars ($>$10 Gyr) show a very large spread, with metallicities ranging between [Fe/H] $= -0.5$ to our metallicity cut-off at $-2.4$.  Note that in all regions there is a non-zero SFR in the most metal-rich two age bins.  This is caused by MATCH attempting to fit the redward extension of the HB (see Fig.\ \ref{cmd}) with an additional, very red RC.  This redward extension is an artefact of the observational dithering pattern in which some (less than $1\%$) of the HB stars have poor $B$-band photometry, hence this detection of very metal-rich, ancient stars is also an artefact, and they are excluded from our star formation history.

Based on spectroscopy of 562 RGB stars in Fornax, \citet{batt06} found two distinct populations: a metal-rich ([Fe/H] $\sim -0.9$) component, and a metal-poor ([Fe/H] $\sim -1.7$) component with a large metallicity spread.  Consistent with observations of other dSphs, they found the metal-rich stars to be more centrally concentrated than the metal-poor population.  To examine the metallicities of the oldest stars, we show histograms of the metallicity distribution of the stars in the oldest age bin in Fig.\ \ref{fig:ancient_feh}.  In Region 1, the metallicity distribution of these ancient stars peaks at [Fe/H] $\simeq -1$, but also contains many of stars with lower metallicities.  The peak at [Fe/H] $\simeq -1$ is also apparent in Region 2, but becomes less strong in Region 3 and is practically absent in Region 4.  This outer bin appears to harbour a broader peak at [Fe/H] $\simeq -1.5$, and a third peak at [Fe/H] $\lesssim -2$.  It should be noted that this third peak is most likely an accumulation of more metal-poor stars outside our metallicity range.  However, the histograms of the inner three regions are consistent with the presence of three peaks at metallicities of [Fe/H] $\simeq -1.0$, $-1.5$, and $\lesssim -2.0$ dex.

\begin{center}
\plotone{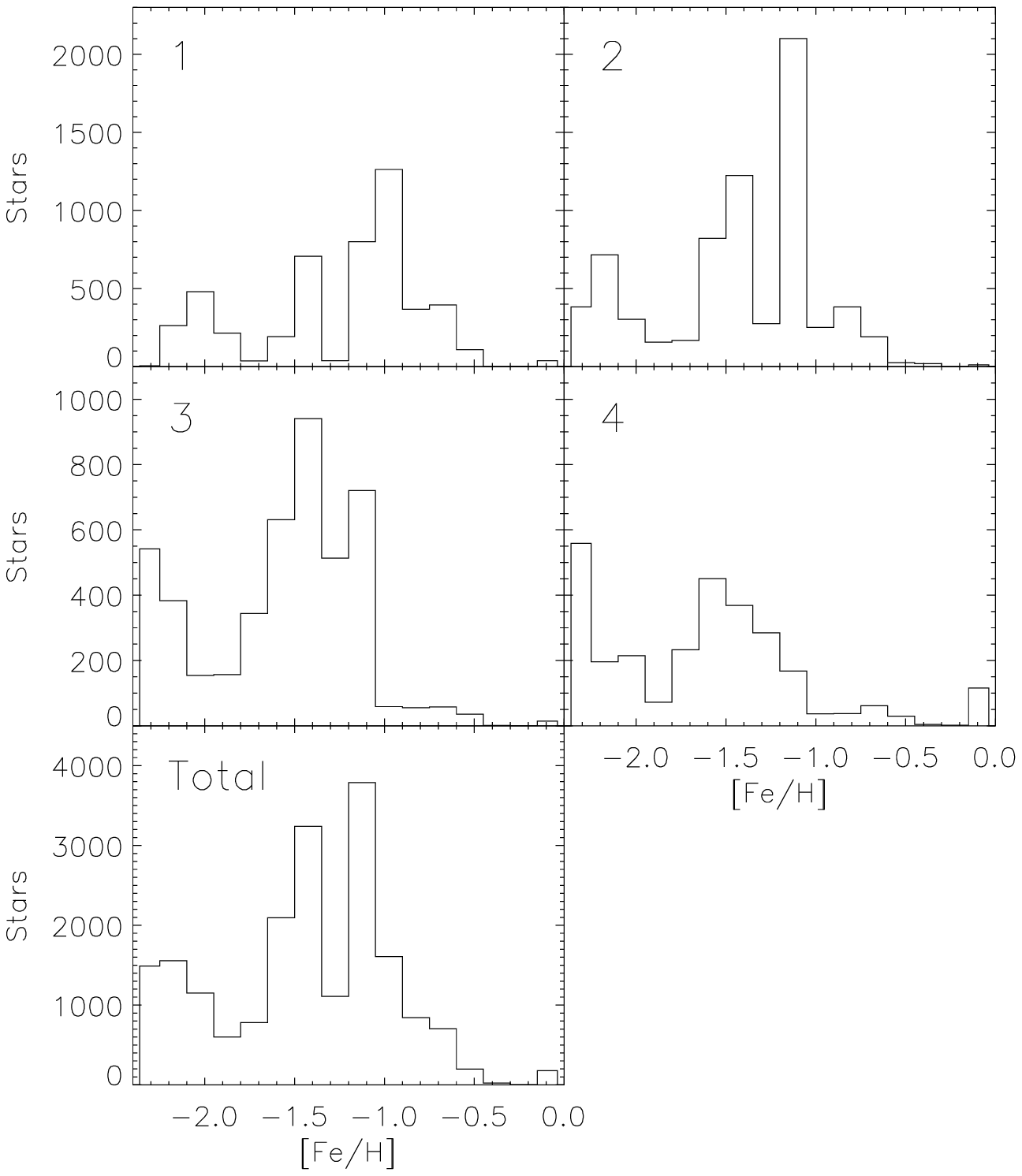} 
\figcaption{ Histograms showing the metallicity distribution of stars in the oldest age bin.  The histograms for the radial bins are labeled 1 through 4 going outwards from the center of Fornax.  Labeled as ``Total'' is the sum of the four radial distributions.  \label{fig:ancient_feh}}
\end{center}

To summarise, we have recovered the metal-rich component (centred at [Fe/H] $\sim -1.0$) of Fornax, however our results support the presence of {\em three} distinct peaks in the metallicity distribution of the ancient stars rather than the two proposed by \citet{batt06}.  To test the statistical significance of the three peaks, we created ten Monte Carlo realisations of the Fornax stellar population using our SFH and age-metallicity relation.  A comparison of the resulting metallicity distributions indicated that, despite some slight differences, the three peaked function was present in every synthetic population.  We therefore argue that the metal-poor component of \citet{batt06} can be sub-divided into two separate populations.  This suggests that Fornax experienced three main star formation events in the period $>$10 Gyr ago, a hypothesis to be tested with further spectroscopic data.

\subsection{The Inner Shell}

In a previous, wide-area survey, \cite{coleman04} noticed a shell-like feature near the center of Fornax in our field 15 (see Fig.\ \ref{obsmap}).  Subsequently, \cite{ols06} confirmed the presence of the feature based on deep photometry obtained with Magellan, and determined an age of 1.4 Gyr and a metallicity of [Fe/H] $\sim -0.7$ for its stars.  Because of the relatively small number of stars in this overdensity and the large contamination by the overall Fornax stars, obtaining a SFH is difficult, as also shown by \cite{ols06}.  Therefore, we opt for constraining the properties of the stars in the feature using a single component (SC) fitting strategy, described in detail in \cite{sdssmatch}.  In short, simple stellar population models with a narrow age and metallicity range are fit to the observed CMD and their goodness-of-fit is compared to that of the best-fitting single component model.

\begin{center}
\plotone{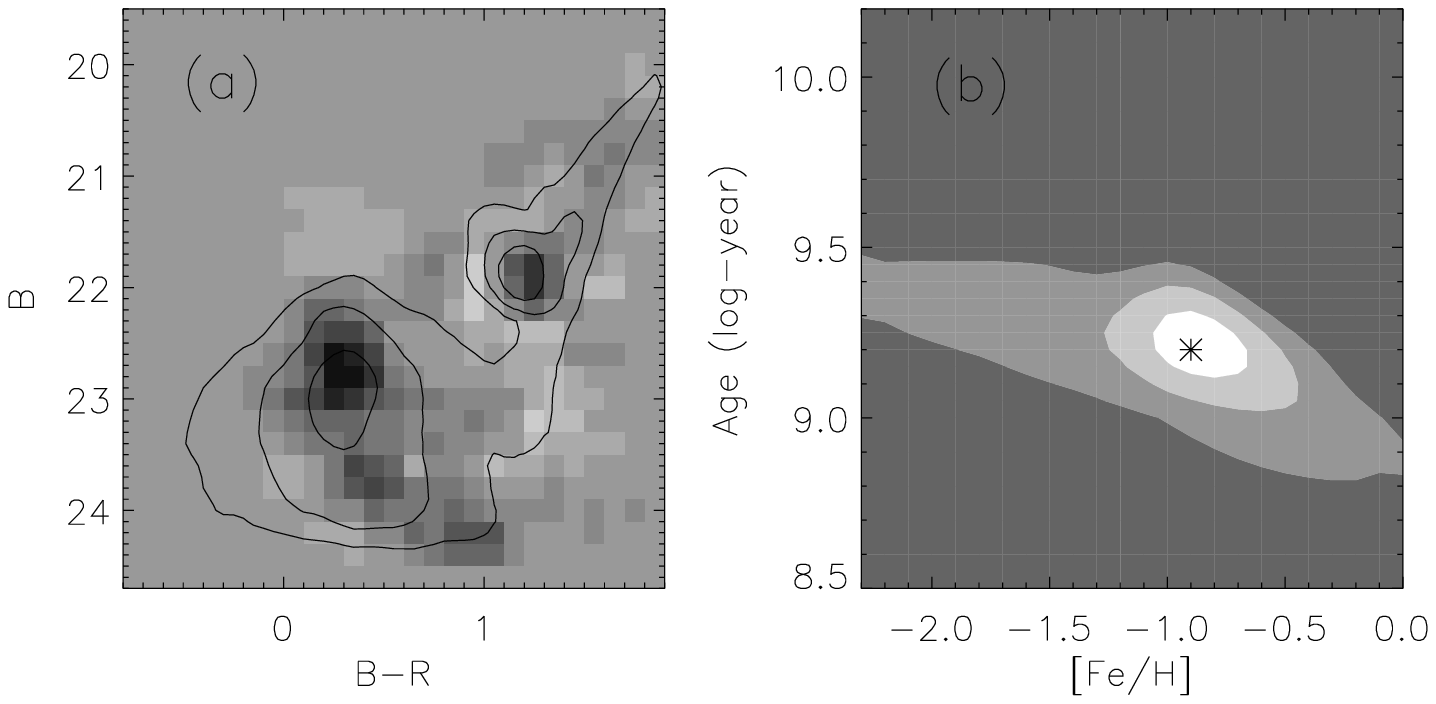}
\figcaption{ Background-subtracted Hess diagram and stellar population constraints for the inner shell.  {\it (a)} Hess diagram of the inner shell feature in grayscale, with darker pixels corresponding to a higher number of stars.  The overplotted contours show the stellar density of the best-fit single-component stellar population.  {\it (b)} Contours showing the regions of the age-metallicity plane where the single-component fits lie within 1, 2, and 3$\sigma$ from the best fit; the best-fit values are indicated with an asterisk.  \label{fig:shell_scfit}}
\end{center}

All stars in a 2\farcm5 wide elliptical annulus containing the inner shell were extracted from the survey.  This was then divided into two regions: the $17^{\circ}$ arc containing the shell, while the remainder of the annulus was used as a control field.  The control field-subtracted Hess diagram of the shell is shown in panel (a) of Fig.\ \ref{fig:shell_scfit}, where the MSTO and RC of the shell population clearly stand out.  Panel (b) of the same figure shows the areas in the age-metallicity plane that lie within 1, 2, and 3$\sigma$ of the best-fit values, indicated with an asterisk.  We find the age of the stars in the shell to be $1.6 \pm 0.4$ Gyr and the metallicity [Fe/H] $= -0.9^{+0.3}_{-0.2}$ dex.  Our SFH results indicate a very low SFR for this age and the average age-metallicity relation predicts a higher metallicity of [Fe/H] $\simeq -0.5$ for stars of the general Fornax population of the same age.  Thus, these results seem consistent with the interpretation of the shell resulting from an accretion event \citep{coleman04,coleman05}, rather than being part of the underlying stellar populations.  This accretion hypothesis will be discussed further in a later publication.

\subsection{Luminosity History}

With the SFH in hand, it is possible to construct the CMD of Fornax as it would have looked at some time in the past.  In this way, the total luminosity of the system can be traced as a function of time.  Combining the SFH fits of all fields, the overall SFR as function of age and metallicity was used to construct artificial CMDs for Fornax at various points in the last 10 Gyr, assuming in all cases a binary fraction of 0.5 and a Salpeter IMF.  By extending the CMDs far enough down the LF, the flux of all stars can be used to calculate the total luminosity.  Although several systematic effects hamper a very precise measurement of $M_V$, the value we obtain for the present day, $M_V \simeq -13.0$, is very close to the literature value of $M_V = -13.1$ \citep{m98}.  The evolution of the V-band luminosity of Fornax during the past 10 Gyr is shown in Fig.\ \ref{fig:lumevol}.  This figure is for illustrative purposes only, and any uncertainties given would be estimates at best.  However, we can see that while Fornax has experienced a general trend towards increasing brightness as more gas is converted to stars, there are significant variations caused by bursts of star formation.  We find that the total luminosity of Fornax has a range of at least 1 magnitude over a Hubble time, hence at some point it was less than $50\%$ of its current brightness.

\begin{center}
\plotone{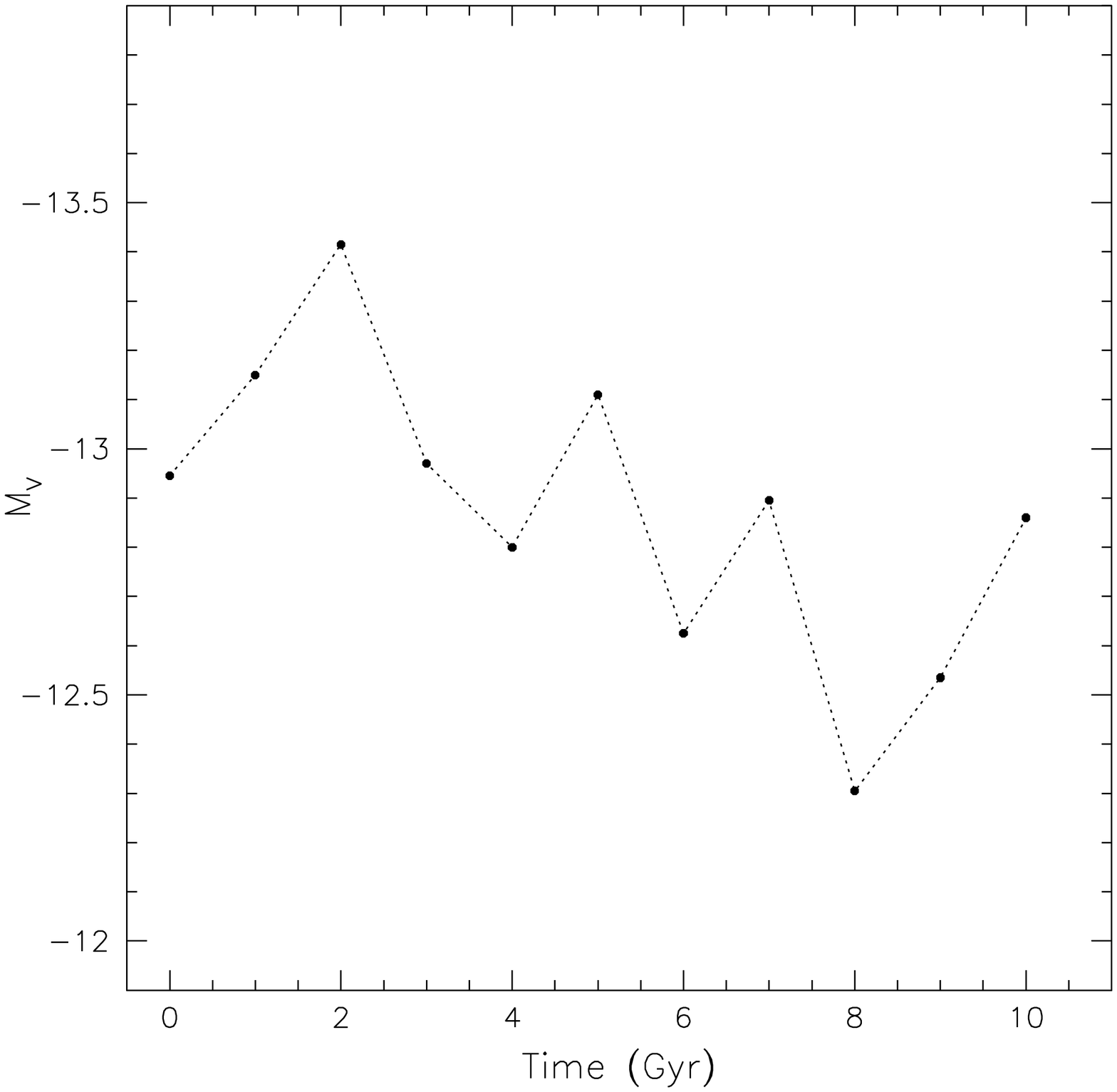}
\figcaption{A schematic diagram of the luminosity evolution of Fornax over the past 10 Gyr.  \label{fig:lumevol}}
\end{center}

\section{Discussion}

The mechanisms of star formation and chemical enrichment in low-mass galaxies are not well understood.  Qualitatively, the process is simple to envision.  Initially, gas collapses at the centre of a dark halo to form the first generation of stars.  The heaviest of these quickly evolve and blowout material, injecting chemicals and energy into the surrounding gas cloud.  This stellar feedback causes the enriched gas cloud to expand and, if the potential well is deep enough, then collapse back to the centre of the dark halo to create a new generation of chemically enriched stars.  Thus, a cyclical process is established in which multiple generations of stars form with a progressive increase in chemical abundance.

However, there are still many unresolved questions regarding dSph star formation.  \citet{helmi06} have shown that, in contrast to the Galactic Halo, the dSphs are conspicuously lacking stars with [Fe/H] $< -3.0$.  This suggests that the gas sourcing the oldest stars in dSphs was pre-enriched.  Given that all dSphs are thought to contain some fraction of ancient stars (i.e.\ with ages $> 13$ Gyr; \citealt{held00}) the initial enrichment must have been an extremely rapid process, via some mechanism which is unclear.

Furthermore, the SFHs for each dSph are remarkably different.  Some dSphs contain purely old stars (ages $> 10$ Gyr) and are characterised by a simple SFH (e.g.\ Draco), whereas others are dominated by intermediate-age stars and have been able to maintain multiple epochs of star formation and chemical enrichment (e.g.\ Fornax).  This is the leading question concerning star formation in dwarf galaxies: why do Draco and Fornax reside at the centres of dark halos with similar masses \citep{walker07}, yet they differ in brightness by a factor of more than ten?  \citet{mayer06} discussed this point, and described two possible scenarios: (i) Either Fornax and Draco initially contained the same amount of gas, hence local effects have allowed Fornax to produce stars with a greater efficiency, or, (ii) Fornax had access to a tenfold greater reservoir of gas, and Draco experienced a similar (but scaled down) initial star formation.  We note that these are extreme scenarios, and something in between is not precluded.

\citet{grebel04} found that the reionization of the universe did not cause the expected reduction in dSph star formation, hence `local effects' are thought to be the dominant factor producing the variety of dSph SFHs.  Differences in the level of tidal distortion, mechanical feedback and gas infall experienced by each system are cited, however the precise nature of these local effects is uncertain (e.g.\ \citealt{dekel86,mayer06}).  In this regard, \citet{ferrara00} have noted that a dSph's dark matter fraction will influence its SFH.  External forces (e.g.\ tidal and ram pressure stripping) can remove blown-out gas from a dSph, however a massive dark halo will allow the satellite to retain its gas.

\subsection{Ancient Stars}

Fornax contains a large number of stars and has a complex star formation history, hence it is an ideal object to compare to simulations.  Firstly, our best fits show that the ancient stars (age $> 10$ Gyr) in this system have a {\em mean} metallicity of [Fe/H] $\approx -1.4$ (Fig.\ \ref{fig:totalsfh}).  This indicates that the first few Gyr contained an intense period of star formation and chemical enrichment.  At some level, this enrichment appears to have occurred throughout the whole body of Fornax: Fig.\ \ref{fig:radialfit} indicates that the oldest stars in every radial bin contain a number of [Fe/H] $\sim -1.0$ stars.  However, we also note a metallicity gradient in this ancient population, such that the central stars display a mean iron abundance approximately 0.3 dex greater than those in the outer regions.  This is consistent with the spectroscopic results of \citet{batt06}.  The three peaks in the metallicity distribution function (Fig.\ \ref{fig:ancient_feh}) are possibly evidence for three main bursts for star formation in the early Universe.  In summary, while our results indicate that the first few Gyrs saw a swift chemical enrichment process in Fornax, they also show that this enrichment was enhanced towards the centre of the system.

In general, our results for the first few Gyr of Fornax are well reproduced by models.  \citet{marcolini06} constructed a 3D hydrodynamic model describing gas dynamics and chemical enrichment in a dwarf galaxy including the contribution of supernovae.  The metalicity distribution function we find for the ancient Fornax stars is similar to that produced by the Marcolini et al.\ model.  Furthermore, they find that stars located towards the centre of a dSph are the product of a more efficient chemical enrichment \citep{marcolini08}, as is seen in Fornax \citep{batt06}.  \citet{salv08} presented a semi-analytic cosmological model following star formation in a dSph galaxy in a Milky Way-type environment, pre-enriched to an abundance of [Fe/H] $\sim -3$.  They showed that a dSph experiences intense star formation in the first few hundred Myr, with multiple cycles of star bursts followed by gas blowout and infall, with accompanying chemical enrichment.  Their metallicity distribution function and mean metallicity are roughly equivalent to those we measured for the old stars in Fornax.  It is clear that simulations are able to accurately reproduce the first epoch of star formation in a dSph environment.

\subsection{Intermediate Age Stars}
\citet{salv08} find a rapidly decreasing star formation rate with time, such that approximately 2.5 Gyr after virialization the rate has fallen well below $10^{-4}$ $M_{\odot}$ yr$^{-1}$.  This corresponds to an age of $\sim$9 Gyr, or the second data point in our global SFH for Fornax (Fig.\ \ref{fig:totalsfh}), where we measure a star formation rate of approximately $3 \times 10^{-3}$ $M_{\odot}$ yr$^{-1}$.  Hence, although the \citet{salv08} model provides an excellent reproduction of star formation in a dSph such as Sculptor, they note that a more complex SFH (such as that seen in Fornax) requires a different set of conditions.

In this context, we find the star formation rate in Fornax to be an approximately constant value\footnote{One should always remember that each datapoint in our SFH represents an {\em average} measurement, and hence will contain hidden complexities.  A useful time frame estimate for a single cycle of star formation and chemical enrichment is $\sim$250 Myr.} of $3 \times 10^{-3}$ $M_{\odot}$ yr$^{-1}$ in the period from 9 to 4 Gyr ago.  This period also witnessed a slow, monotonic increase in iron abundance.  However, star formation in Fornax experienced a sudden increase approximately 3 to 4 Gyr ago, jumping threefold to $\sim$$10^{-2}$ $M_{\odot}$ yr$^{-1}$ with an accompying spike in chemical enrichment.  These results all agree with the Ca {\sc ii} triplet results of \citet{pont04}.  Our results also suggest that this epoch of star formation was relatively short-lived, lasting $1-2$ Gyr, and was confined to the central $\sim$$0.5r_t$ (1500 pc) of Fornax (Fig.\ \ref{fig:radialsfh}).

There are a variety of explanations for this continued star formation.  \citet{salv08} state that a refinement of the reionization criterion would allow massive dSphs with a lower initial gas-to-dark ratio in their models, thereby leading to less efficient mechanical feedback and more regular star formation activity.  This could explain the steady star formation seen in the period from 9 to 4 Gyr ago, however it cannot account for the subsequent burst.  As an alternative scenario, Fornax may have experienced an injection of new gas to fuel this next generation of stars.  We have already presented evidence for a merger in Fornax, proposing that a gas-rich dwarf galaxy merged with this system to fuel strong star formation activity \citep{coleman04,coleman05}.  However, the timing is problematic: the original scenario requires that this merger occured approximately 2 Gyr ago, or 2 Gyr {\em after} the suddent burst of star formation found here.  Furthermore, our chemical enrichment history shows that the new burst of stars was accompanied by a sudden increase in abundance, hence this would imply that, prior to star formation, the metal abundance of the gas was at least that of Fornax and therfore unlikely to be of foreign origin.  We would therefore argue that the 4 Gyr burst was fueled by gas originating in Fornax.

This scenario requires gas blown away by star formation to have resided in the outer regions of Fornax for at least 5 Gyr before collapsing back to the dSph.  Gas expelled to the outer regions (or even the halo) of a satellite system is generally expected to be removed by ram pressure stripping and tidal distortion.  However, \citet{blitz00} presented evidence that H{\sc i} clouds exist in many Local Group dSphs, situated up to 10 kpc (or, approximately three tidal radii in the case of Fornax) from the centre of the system.  This is supported by the strong evidence for H{\sc i} associated with Sculptor \citep{carignan98,bouchard03}.  Indeed, \citet{bouchard06} discovered an H{\sc i} cloud located to the North of Fornax.  Whether this is associated with the dSph is not certain (the radial velocity of Fornax is inconveniently close to that of the Milky Way in this direction), however \citet{bouchard06} state a minimum mass of $1.5 \times 10^5$ $M_{\odot}$ at the distance of Fornax.  \citet{mayer05} noted that an object following an orbit with a low ellipticity (such as Fornax; \citealt{piatek07}) will be better able to retain its gas.  It therefore seems possible that the massive, extended dark halo of Fornax could have allowed gas in the outer regions to remain bound to the system for an extended period.

An external influence on star formation in a satellite system are tidal interactions with the larger host galaxy \citep{mayer06}.  Tidal forces experienced by a satellite system as it orbits its host can induce burts of star formation \citep{barton00} as the interstellar gas clouds are compressed \citep{mihos96}.  Using {\em HST} images over a four year epoch, \citet{piatek07} measured the proper motion of Fornax and derived an orbital period of $3.2^{+1.4}_{-0.7}$ Gyr with an eccentricity of $e=0.13^{+0.25}_{-0.02}$.  Tidal forces scale as $R^{-3}$, hence this orbit implies that Fornax will experience a distortion force change of at least $50\%$ as it moves from pericentre to apocentre.  A pericentric passage approximately 4 Gyr ago is possible within the current solution (we are grateful to S.\ Piatek for sharing his orbital code), however the uncertainties are currently too large for a fair comparison between orbit and SFH.

\subsection{Recent Activity}

Finally, we examine the recent activity in Fornax.  Following the 4 Gyr burst, there was another star forming event $400-600$ Myr ago, and a more recent event approximately 100 Myr ago (Fig.\ \ref{fig:totalsfh}).  These 100 Myr old stars have been previously noted by \citet{stetson98} and \citet{saviane00}, and are the youngest stars yet observed in a dSph.  However, as a new result, we have also detected a small number of younger stars.  Table \ref{tab:totalsfh} indicates that in the period $10-100$ Myr ago, $\sim$1500 $M_{\odot}$ were formed in Fornax.  This is minute compared to Fornax as a whole (see Fig.\ \ref{masshistory}), yet it is tentative evidence that {\em Fornax may have been forming stars almost to the present day}.  We ignore the abundance measurements for this youngest population, given that they are based on only a few stars in each age bin and have significant errors.

\begin{center}
\plotone{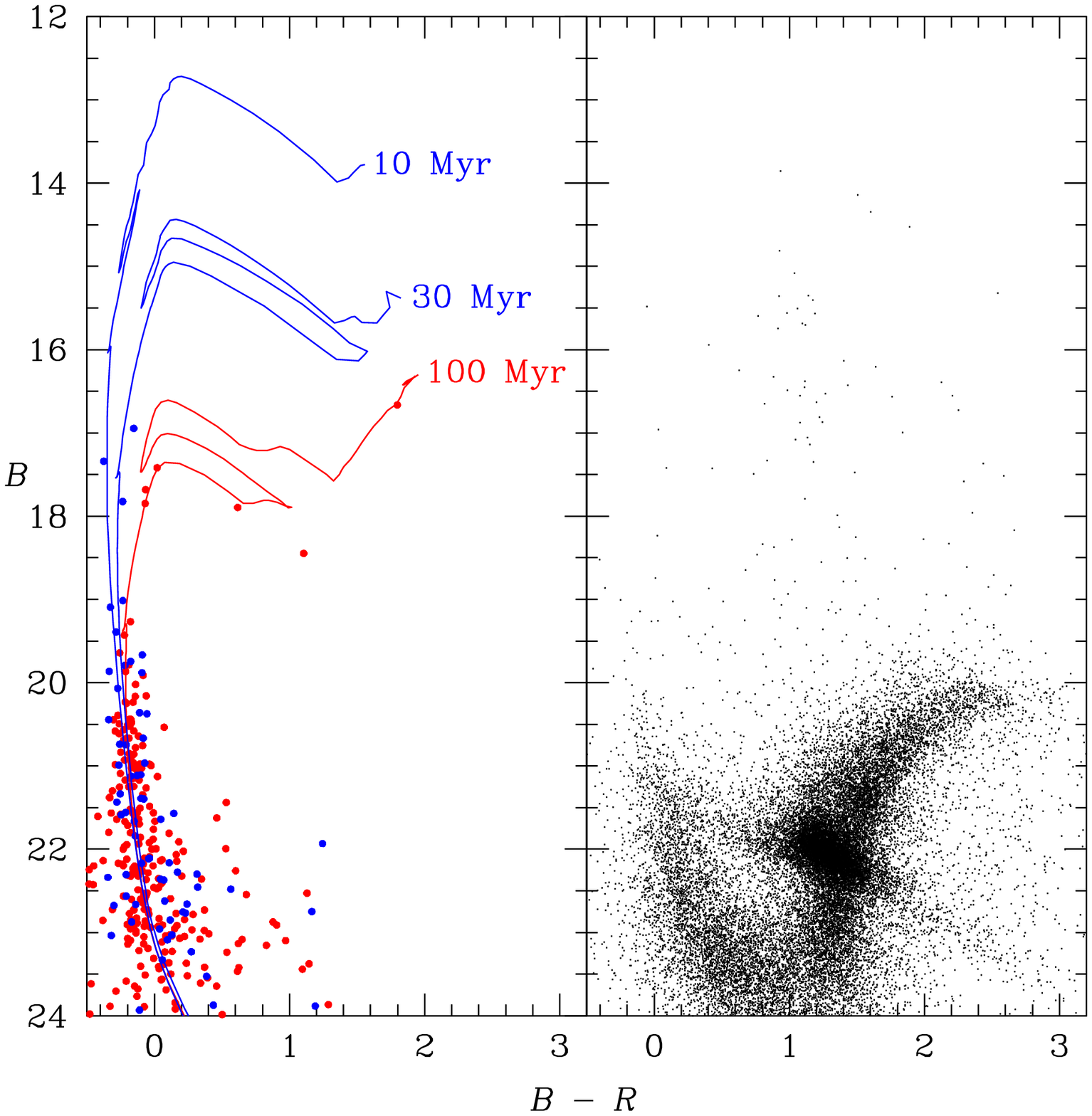}
\figcaption{Artificial CMD of the $\le 100$ Myr stellar populations compared to that of the Fornax core region.  The left panel shows a typical colour-magnitude distribution for the young stars in Fornax as predicted by our star formation history.  The red points trace a 100 Myr population (the $7.95 - 8.10$ $\log{\mbox{(yr)}}$ age bin in Table \ref{tab:totalsfh}) while the blue points represent the $<$100 Myr stars.  The right panel shows the CMD of the core region of Fornax. \label{youngcmd}}
\end{center}

To demonstrate the difficulty in detecting this small, age $< 100$ Myr population, we show an artificial CMD of the Fornax young stars in Fig.\ \ref{youngcmd}.  The left panel is a `typical' prediction\footnote{This prediction is susceptible to a variety of assumptions such as the IMF and binary fraction  Here we have used the same recipe as described above; a Salpeter IMF with a binary fraction of 0.5.  We constructed a large number of these artificial young populations and chose one which is most indicative of the class.} for the young stars based on our SFH.  Extinction, photometric errors and completeness for the survey have been included.  Note that the distribution of the blue points (age $< 100$ Myr) is almost identical to that of the red points (age $= 100$ Myr), thus these populations are degenerate and it is not possible to distinguish between them based only on photometry.  However, it shows that such a very young population could easily have evaded detection up to now.  Given the small number of young stars predicted by our SFH, and the difficulty in separating them from slightly older stars, we classify this as tentative evidence for an ultra-young population in Fornax.  Nonetheless, we expect very little foreground contamination blueward of $B - R = 0.4$, hence the brightest blue stars shown in the right panel would argue that Fornax did indeed experience star formation less than 100 Myr ago.  High resolution spectra of these young stars is required to accurately determine their age.  Lithium is an important age diagnostic for late-type stars as it is easily destroyed in stellar interiors, hence a follow-up survey could target the resonance doublet of Li {\sc i} at 6708 \AA \ (e.g.\ \citealt{montes01}) in these young stars to demarcate the age of the most recent star forming event in Fornax.

\section{Summary and Conclusions}

Based on two-filter photometry to a magnitude of $B \sim 23$, we used a CMD-fitting technique to derive the star formation history for the Fornax dSph.  All dSphs contain some number of ancient stars, however these systems are known to display a wide variety of SFHs.  Fornax formed a significant number of its stars in the early Universe (age $> 10$ Gyr) and subsequently experienced a constant star formation rate.  This behaviour can be reproduced by the simulations.  However, in the period $3-4$ Gyr ago, Fornax experienced a sudden burst of star formation, approximately three times the rate for the previous 5 Gyr.  The cause of this activity is unclear.  The star formation rate has been decreasing ever since, with smaller events $400-600$ Myr and 100 Myr ago.  We also find tentative evidence for a small number of stars (total mass $\sim$$1500 M_{\odot}$) which have formed in the past 100 Myr.  Fornax contains the most recent star formation activity of any Local Group dSph.

Strong radial gradients in the SFH are also evident.  As noted by previous authors \citep{stetson98,saviane00}, recent star formation has been concentrated towards the centre of Fornax.  This trend is seen in other dSphs \citep{harbeck01}.  Furthermore, we have found that chemical enrichment was more efficient at the dSph's centre.  Even the oldest stars display a metallicity gradient, such that the inner stars have a mean iron abundance approximately $0.3$ dex greater than the outer stars.  Indeed, the first few Gyr in Fornax appear to have been a time of intense star formation and chemical enrichment: the age $> 10$ Gyr stars display a mean abundance of [Fe/H] $=-1.4$.  The oldest stars in Fornax also show three peaks in the metallicity distribution, possibly evidence for three main bursts of star formation in the period $>$10 Gyr ago.  We find the metallicity to have increased monotonically in the period $9-4$ Gyr ago, and then experienced a sharp increase in conjunction with the intense burst of star formation described above.

Thus, while the first few Gyr of star formation in Fornax can be reproduced by models, the cause of the burst 4 Gyr ago is unclear.  We have previously proposed that a gas-rich dwarf galaxy merged with Fornax approximately 2 Gyr ago to produce sub-structure \citep{coleman04,coleman05}, however we cannot reconcile the timing of this event with the observed peak in the SFH.  Therefore, we suggest that gas enriched and blown out by earlier star forming events H{\sc i} clouds settled in the outer regions of Fornax, and then re-collapsed to fuel an intense period of star formation at the centre of the dSph.  This event may have been caused by tidal interactions with the Milky Way during a pericentric passage.

\acknowledgments
The authors thank the anonymous referee for their comments which have improved the manuscript.  The authors also thank S.\ Piatek and C.\ Pryor for their helpful advice regarding the orbit of Fornax in different potentials, and we acknowledge the use of their orbital algorithms.  We are indebted to H.-W.\ Rix for his helpful discussions, and the MPIA observing team who obtained the MPG/ESO 2.2m data described in this paper.

{\em Facilities:} Max Plank:2.2m

\clearpage


\begin{table}
\begin{center}
\caption{Fornax Star Formation History}
\label{tab:totalsfh}
\vspace{0.2cm}
\begin{tabular}{lccc}
\tableline
\tableline
Age Bin & SFR & [Fe/H] & Mass \\
(log-yr) & ($\times$10$^{-3} M_{\odot}$ yr$^{-1}$) &  & ($10^3 M_{\odot}$) \\
\tableline
$7.05-7.20$ & $0.16 \pm 0.09$ & $-2.2 \pm 0.9$ & $0.73 \pm 0.40$ \\
$7.20-7.35$ & $0.02 \pm 0.01$ & $-1.9 \pm 0.5$ & $0.16 \pm 0.08$ \\
$7.35-7.50$ & $0.09 \pm 0.04$ & $-1.6 \pm 0.6$ & $0.81 \pm 0.29$ \\
$7.50-7.65$ & $0.0$ & --- & $0.0$ \\
$7.65-7.80$ & $0.0$ & --- & $0.0$ \\
$7.80-7.95$ & $0.0$ & --- & $0.0$ \\
$7.95-8.10$ & $0.15 \pm 0.07$ & $-1.9 \pm 1.1$ & $5.5 \pm 2.6$ \\
$8.10-8.25$ & $0.0$ & --- & $0.0$ \\
$8.25-8.40$ & $0.04 \pm 0.03$ & $-1.2 \pm 0.9$ & $3.0 \pm 1.9$ \\
$8.40-8.55$ & $0.11 \pm 0.04$ & $-0.6 \pm 0.7$ & $11 \pm 4$ \\
$8.55-8.70$ & $0.24 \pm 0.09$ & $-0.8 \pm 0.5$ & $35 \pm 12$ \\
$8.70-8.85$ & $0.75 \pm 0.15$ & $-0.4 \pm 0.3$ & $154 \pm 30$ \\
$8.85-9.00$ & $0.14 \pm 0.06$ & $-0.2 \pm 0.6$ & $42 \pm 15$ \\
$9.00-9.15$ & $0.12 \pm 0.06$ & $-0.9 \pm 0.9$ & $51 \pm 23$ \\
$9.15-9.30$ & $0.17 \pm 0.08$ & $-1.1 \pm 0.8$ & $98 \pm 46$ \\
$9.30-9.45$ & $1.50 \pm 0.63$ & $-0.5 \pm 0.6$ & $1240 \pm 514$ \\
$9.45-9.60$ & $9.4 \pm 1.9$ & $-0.9 \pm 0.4$ & $10900 \pm 2200$ \\
$9.60-9.75$ & $3.0 \pm 1.5$ & $-1.1 \pm 0.3$ & $5000 \pm 2470$ \\
$9.75-9.90$ & $3.6 \pm 1.7$ & $-1.3 \pm 0.3$ & $8400 \pm 3880$ \\
$9.90-10.05$ & $3.2 \pm 1.1$ & $-1.3 \pm 0.4$ & $10500 \pm 3440$ \\
$10.05-10.20$ & $5.4 \pm 1.0$ & $-1.4 \pm 0.3$ & $25000 \pm 4610$ \\
\tableline
\end{tabular}
\end{center}
\end{table}


\clearpage


\end{document}